\documentclass{aa}
\usepackage{graphicx}
\usepackage{txfonts}
\newcommand{\rg}{r_{\mathrm{g}}}
\newcommand{\rms}{r_{\rm ms}}

\begin{document}
\title{The relativistic shift of narrow spectral features from black-hole accretion discs}
\author{T.~Pech\'a\v{c}ek,\inst{\!1} M.~Dov\v{c}iak,\inst{\!2} V.~Karas,\inst{\!2,1} \and G.~Matt\inst{3}}
\institute{$^1$~Charles University, Faculty of Mathematics and Physics, V~Hole\v{s}ovi\v{c}k\'ach~2, CZ-180\,00~Prague, Czech Republic\\
 $^2$~Astronomical Institute, Academy of Sciences, Bo\v{c}n\'{\i}~II, CZ-141\,31~Prague, Czech Republic\\
 $^3$~Dipartimento di Fisica, Universit\`a degli Studi ``Roma Tre'', Via della Vasca Navale 84, I-00146~Roma, Italy}

\authorrunning{T.~Pech\'a\v{c}ek et al.}
\titlerunning{Relativistic shift of narrow spectral features}
\date{Received 17 May 2005; Accepted 14 June 2005}
\abstract{Transient spectral features have been discovered in
the X-ray spectra of Active Galactic Nuclei, mostly in the 5--7~keV
energy range. Several interpretations were proposed for the origin of
these features. We examined a model of Doppler boosted blue horns of 
the iron line originating from a spot in a black hole accretion 
disc, taking into account different approximations of general relativistic 
light rays
and the resulting shift of energy of photons. We provide a practical
formula for the blue horn energy of an intrinsically narrow line and
assess its accuracy by comparing the approximation against an exact
value, predicted under the assumption of a planar accretion disc. The
most accurate approximation provides excellent agreement with the spot 
orbital radius down to the marginally stable orbit of a
non-rotating black hole.
}
\maketitle

\section{Introduction} X-rays are the main source of information about
the innermost parts of accretion discs in active galactic nuclei (AGN)
and Galactic black hole systems. The iron K$\alpha$ line is the most
prominent spectral feature, occurring at a rest energy $E_0$ ranging
from $\sim6.4$~keV (neutral iron) to $\sim6.96$~keV (H--like iron). The
natural width of these lines is only a few eV (Thompson et al.\
\cite{tho01}). They are  thought to originate as a result of
illumination of an accretion disc, which is formed by relatively cold
material, by high-energy continuum photons from a corona or a jet above
the disc (Fabian et al.\ \cite{fab00}; Reynolds \& Nowak \cite{rey03}).
The K$\alpha$ lines are multiplets. The neutral iron
line is a doublet with energies $E_{\mathrm{K}\alpha_1}=6.404$~keV,
$E_{\mathrm{K}\alpha_2}=6.391$~keV. The small energy difference makes
the doublet unresolved with present spectrometers, though it may be
resolvable by ASTRO-E2 in the near future. Another prominent spectral
feature, namely the K$\beta$ line, arises at an energy of $\sim7.07$~keV
and the iron edge occurs at $7.1$~keV. Special and general relativity
effects strongly modify the appearance of all these features. For
instance, the continuum becomes broader with increasing disc inclination
due to enhanced Doppler shifts. The photoelectric edge is blurred into
broad troughs.  X-ray spectroscopy can then be employed  to reveal the
effects of general relativity influencing the observed energy and
thereby shaping the spectral lines and the continuum slope.

Fast circulation of gas is the dominant source of line broadening and
forms, together with substantial gravitational redshift, the observed
profile. A double-horn shape is 
expected in the simplest case of a geometrically thin, Keplerian disc
(Fabian et al.\ \cite{fab89}), and the discovery of this
kind of profile has been indeed reported in several objects, the
best-known examples being the Seyfert galaxy MCG--6-30-15 (Tanaka et al.\
\cite{tan95}) and the microquasar GRS 1915+105 (Martocchia et al.\
\cite{mar02}), where the line shows equivalent widths reaching
$\sim200$~eV. It has been argued that fine substructures of this broad
feature could be used to constrain the inclination angle of the
source, radial emissivity in the disc plane, and even the angular
momentum of the central black hole (see Martocchia et al.\ \cite{mar00};
Beckwith \& Done \cite{bec04}, and references cited therein).
However, very broad ``classical'' double-horn relativistic lines are
often absent in AGN (e.g.\ Bianchi et al.\ \cite{bia04}), and it is
important to seek additional evidence.

In recent years, {\em narrow} emission features have been reported in
the X-ray spectra of several AGNs: see Turner et al.\ (\cite{tur02},
\cite{tur04}, \cite{tur05}); Guainazzi (\cite{gua03}); Yaqoob et al.\
(\cite{yaq03}); Bianchi et al.\ (\cite{bia04});  Dov\v{c}iak et al.\
(\cite{dov04}); Gallo et al.\ (\cite{gal04}); McKernan
\& Yaqoob (\cite{mck04}); Porquet et al.\ (\cite{por04}); Della Ceca et al.\
(\cite{cec05}). This topical discovery renewed the interest in accretion
disc hot spots. The narrow lines we have on mind are typically found in
the $5$--$7$~keV energy range, i.e.\ mostly on the redshifted side of the iron
line rest energy. A tentative explanation for these puzzling features,
which we adopt in this paper, is in terms of iron emission,
which originates in a localized area of the disc
($\delta{r}\lesssim\rg$). According to this scheme the
spectral features are produced by transient magnetic flares
briefly illuminating the disc and producing the iron line by
fluorescence in the surface layer (Haardt et al.\
\cite{har94}; Poutanen \& Fabian \cite{pou99}; Czerny et al.\ \cite{cze04}).

The idea of revealing the signatures of black holes via spectroscopy of
orbiting spots has great potential
and a long history (Cunningham \& Bardeen \cite{cun73}). Theoretical line
profiles have been studied including specific effects  of general
relativity: the frame-dragging (Kojima \cite{koj91}; Laor
\cite{lao91}; Bromley et al.\
\cite{bro97}), extreme light-bending and multiple images 
(Matt et al.\ \cite{mat93}; Viergutz \cite{vie93}; Bao et al.\
\cite{bao94};  Zakharov \& Repin \cite{zak03}; Beckwith \& Done
\cite{bec05}), non-axisymmetric (spiral) waves in accretion discs (Karas
et al.\ \cite{kar01}; Hartnoll \& Blackman \cite{har02}), and self-gravity
(Karas et al.\ \cite{kar95}; Usui et al.\ \cite{usu98}).
Various applications have been proposed for the general-relativity 
energy shift of spectral features from co-rotating spots; in particular,
it was suggested that the range of energy fluctuations could constrain 
the source geometry and discrimate between planar and thick discs 
(Bao \& Stuchl\'{\i}k \cite{bao93}). 

In this paper we examine extremal values (maximum and minimum) of
observed energy of a line from an orbiting spot. The main purpose of
this study is to provide a simple, practical and accurate formula that
will be useful in analyzing the redshifted narrow
lines. We neglect
light bending, because its influence is important for only a relatively
minor fraction of photons on special rays, i.e.\ those passing near
caustics. This approximation was assessed by Gerbal \& Pelat
(\cite{ger81}), but these authors did not aim to provide an explicit
formula that would be convenient for practical purposes. 
Our approach is similar to Zhang \& Bao (\cite{zha91}), who also
ignored light bending. We improve their treatment of light aberration
and this allows us to express the redshift function (defined as a ratio
of observed to emitted energy; see eq.~(\ref{eq:gfac}) below) in a way 
that is comparably 
simple to but more accurate than their eq.~(2.9). As a further step, we include 
light bending in terms of Beloborodov's (\cite{bel02}) approximation and
reach fractional accuracies of a few percent or better for the
redshift function; the extremal values are given with relative error
better than $0.1$\%. Thus the resulting shift of energy is estimated at least 
an order of magnitude more precisely than before, and the error of 
its extremal values is negligible in most practical cases.

In the next section we summarize the main results of this paper. In
particular, in subsection~\ref{sec:summary} we specify our assumptions 
and give a prescription for calculating the observed energy range
spanned by the spectral line (i.e. extremal values of $g$-factor). It
will be of interest to a reader  seeking a straightforward approach
to practical problems of data fitting. Then, in
subsec.~\ref{sec:details}, we describe the individual steps of the 
derivation, including a
comparison with previous results of other authors. In
Sec.~\ref{sec:results}, in a brief re-discussion of several AGNs we
illustrate the interpretation of narrow spectral  features in terms of
the blue horn produced by an orbiting spot. The results presented
here are limited to static black holes and geometrically thin, Keplerian
discs. Even with these limitations, they may be useful in interpreting
the present data and deriving disc parameters. As it will be
shown in section~\ref{sec:results}, the narrow features detected so far do
not require spinning black holes.

\section{Calculation of the redshift}
\subsection{Summary}
\label{sec:summary}
For the gravitational field of the system we assume a non-rotating
black hole and so adopt the Schwarzschild spacetime
metric (Misner et al.\ \cite{mis72}),
\begin{equation}
\mathrm{d}s^2=-\frac{r-2}{r}\;\mathrm{d}t^2+\frac{r}{r-2}\;\mathrm{d}r^2+r^2\;
\left(\mathrm{d}\theta^2+\sin^2{\!\theta}\ \mathrm{d}\phi^2\right),
\label{eq:ds}
\end{equation}
where the radius $r$ is evaluated in units of gravitational radius
$\rg\equiv{GM}c^{-2}\dot{=}1.5\times10^5(M/M_{\sun})$~cm, with $M$ being
the mass of the black hole. In eq.~(\ref{eq:ds}) as well as everywhere in this
text, geometrized units with $c=G=1$ are used; distances are thus made 
dimensionless by scaling them with $M$.
Notice that the accretion disc own mass can be neglected, i.e.\
$M_\mathrm{d}{\ll}M$, because we constrain ourselves to the innermost
regions of the flow where the disc is non-self-gravitating (in terms of
Toomre's criterion).

The disc is assumed to be Keplerian and extending down to the
marginally stable orbit, $\rms=6$, below which it ceases to exist.
An observer is 
located at infinity on the $x$ axis (i.e.\ $\phi=0^\circ$).
By approximating trajectories of light according to Beloborodov 
(\cite{bel02}) and evaluating the effect of aberration as detailed
in subsec.~\ref{sec:details}, we obtain the extremal values of the 
energy shift in the form
\begin{equation}
\label{beloborodov_mm}
g_{\mbox{\tiny B}}^\pm = \frac{r^{1/2}\,\left(r-3\right)^{1/2}}{r+\left[r-2+
            4\left(1+\cos{\phi_\pm}\sin{\theta_\mathrm{o}}\right)^{-1}\right]^{1/2}
            \sin{\phi_\pm}\sin{\theta_\mathrm{o}}}.
\end{equation}
These extremal values occur for the azimuthal polar coordinate 
$\phi\equiv\phi_{\pm}$, defined by
\begin{eqnarray}
\sin{\phi_{\pm}} & = & \mp\left(1-\cos^2{\phi_{\pm}}\right)^{1/2}, \\
\cos{\phi_{\pm}} & = & {f_1}^{-1}\left[{f_2}^{1/3}+(r-4)^2 
 {f_2}^{-1/3}-2(r-1)\right]
\end{eqnarray} 
($\phi=90^\circ$ corresponds to the maximally receding part of the disc).
The functions $f_1(r,\theta_\mathrm{o})$ and $f_2(r,\theta_\mathrm{o})$ 
are
\begin{eqnarray}
\label{f1}
f_1(r,\theta_\mathrm{o}) & = & 3\,(r-2)\,\sin{\theta_\mathrm{o}},\\
\label{f2}
f_2(r,\theta_\mathrm{o}) & = & \left(r-4\right)^3+{f_3}^2+
                         \left[2\left(r-4\right)^3+{f_3}^2\right]^{1/2}f_3,
\end{eqnarray}
with $f_3(r,\theta_\mathrm{o})$ being 
\begin{equation}
\label{f3}
f_3(r,\theta_\mathrm{o}) = 3\sqrt{3}\;(r-2)\,\cos{\theta_\mathrm{o}}.
\end{equation}

Equation (\ref{beloborodov_mm}) can be used to determine the range of 
energy spanned by a line without the need
to evaluate special functions (elliptic integrals and their inverse in
case of Schwarzschild black hole; e.g.\ Chandrasekhar \cite{cha92};
\v{C}ade\v{z} et al.\ \cite{cad03}) or to perform numerical ray-tracing 
in the curved spacetime. Instead, purely algebraic formulae are involved, 
allowing us to give extremal values of redshift
as functions of the spot radius
$r$ and observer's inclination angle
$\theta_\mathrm{o}$. Below we estimate the accuracy of our
approximation and find it very precise even at large values of
inclination (almost edge-on, i.e.\ $\theta_\mathrm{o}\sim89^\circ$) and
small radii (down to the marginally stable orbit,
$r\sim\rms$). We find much better agreement between approximate and exact
values than what can be achieved with the Zhang \& Bao (\cite{zha91})
approach.

\begin{figure*}
\begin{center}
\includegraphics[width=0.47\textwidth]{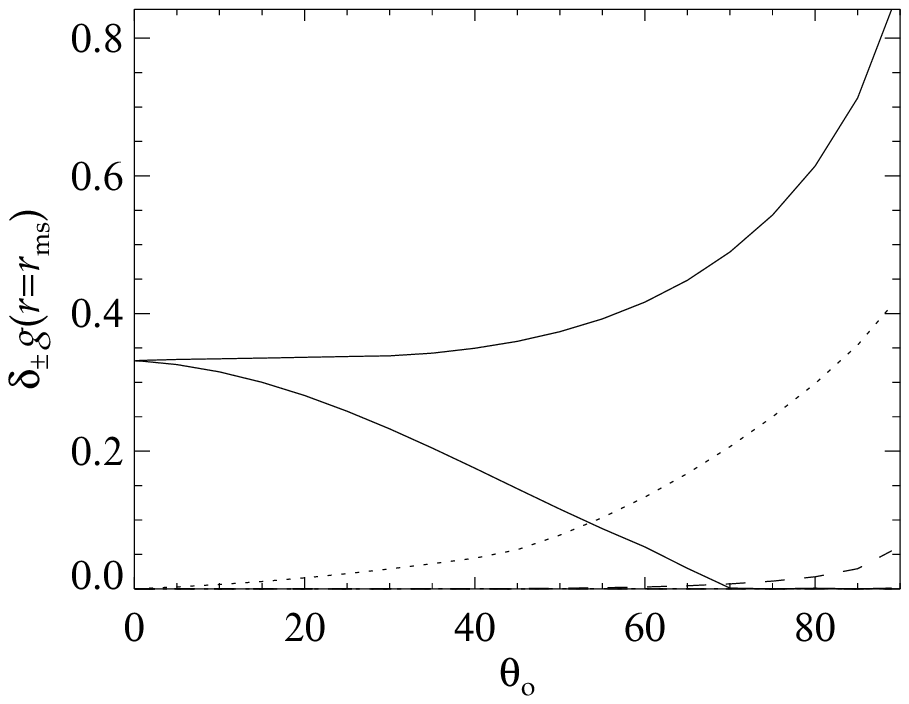} 
\hfill
\includegraphics[width=0.47\textwidth]{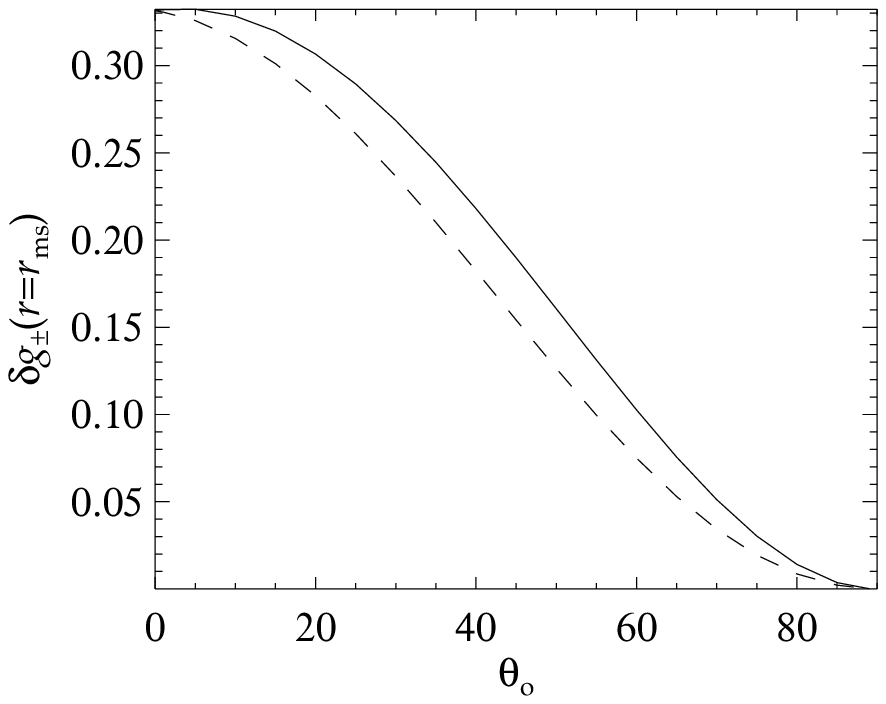}
\includegraphics[width=0.47\textwidth]{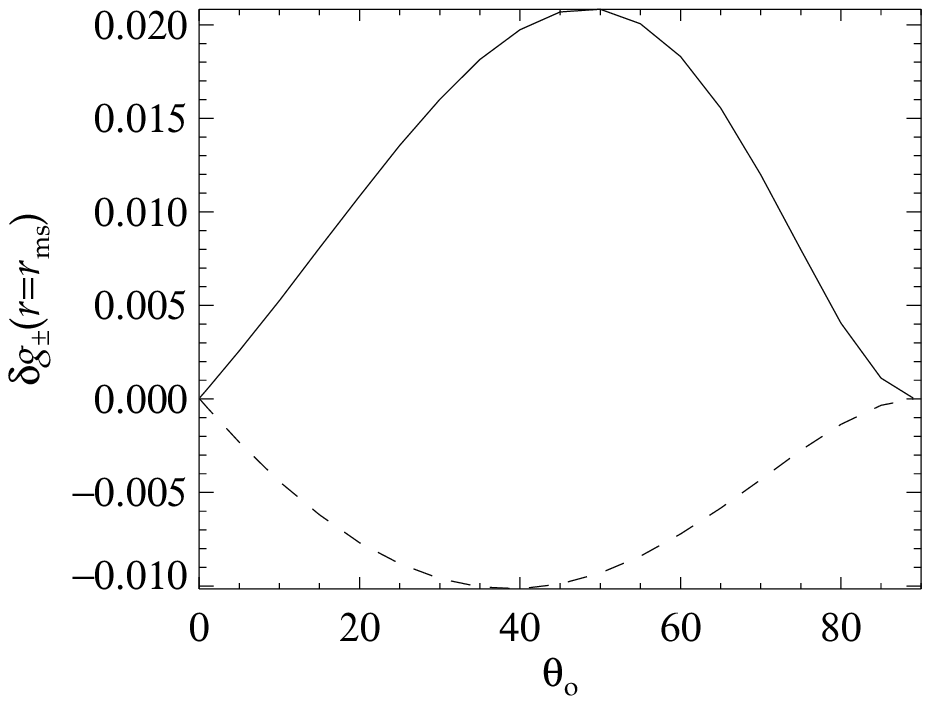} 
\hfill
\includegraphics[width=0.47\textwidth]{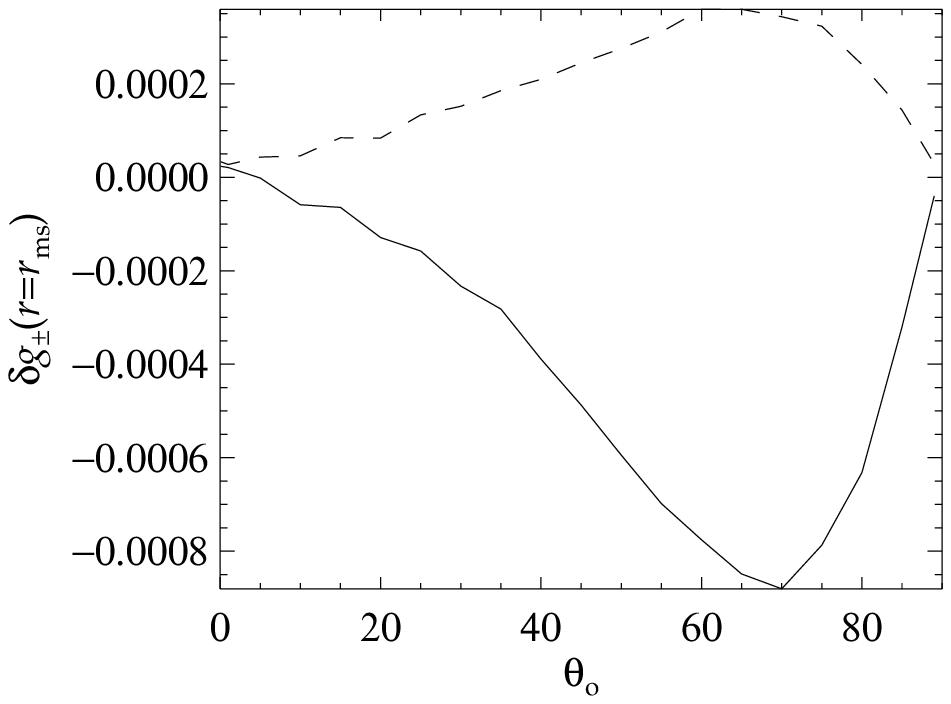}
\caption{Accuracy assessment for different approximation formulae. 
(i)~Top row -- left panel: the dependence of the maximum and minimum relative
error $\delta_\pm g(r,\theta_\mathrm{o})$ of the $g$-factor on observer's
inclination angle $\theta_\mathrm{o}$ for a narrow ring, $r=r_{\rm ms}$. 
Solid lines refer to eq.~(\ref{bao}) (the upper
one is for maximum, the lower one is for minimum); the dotted line
refers to eq.~(\ref{sl}); the dashed line is for
eq.~(\ref{beloborodov}). The approximation formula (\ref{bao}) is the worst
of the three cases in the sense that it has a significant non-zero value
of the minimum error. This is due to  neglecting aberration. The
relative error  $\delta g = |g_{\rm approx}-g|/g$ is
determined with respect to exact (numerically computed) values. (ii)~Top row
-- right panel: the relative error $\delta g_\pm(r,\theta_\mathrm{o})$ of 
the extremal values of $g$-factor for a ring $r=r_{\rm ms}$ as a function
of observer's inclination angle. The relative error 
$\delta g_\pm = (g_{\rm approx.}^\pm-g_\pm)/g_\pm$ of the extremal 
values is again
determined with  respect to exact values from equation of geodesic. The
solid/dashed lines are for maximum/minimum of $g$-factor; both lines are
for the approximation (\ref{bao_mm}). (iii)~Bottom row -- left panel:
the same as top right but with the approximation (\ref{sl_mm}). (iv)~Bottom
row -- right panel: the same as top right but for the approximation
(\ref{beloborodov_mm}).}
\label{fig1}
\end{center}
\end{figure*}

\begin{figure*}
\begin{center}
\includegraphics[width=0.49\textwidth]{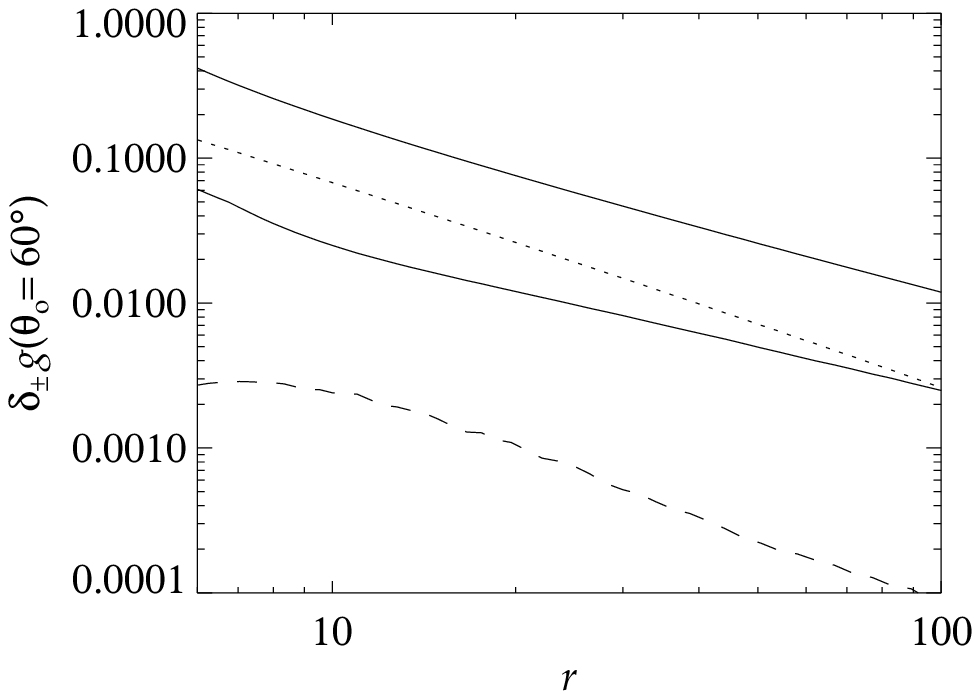}
\includegraphics[width=0.49\textwidth]{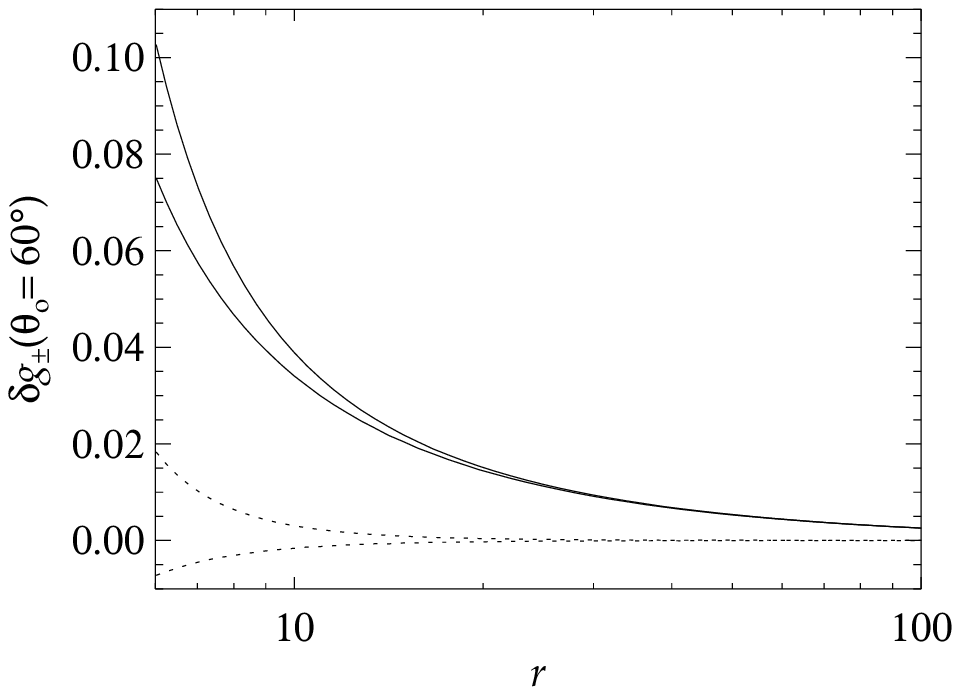}
\caption{Left: the dependence of the maximum and minimum relative error 
$\delta_\pm g(r,\theta_\mathrm{o})$ of
$g$-factor on the radius for inclination 
angle $\theta_\mathrm{o}=60^\circ$, in logarithmic scale. 
Solid lines are for the approximation (\ref{bao}) 
(upper for the maximum, lower for the minimum), dotted line is for the approximation 
(\ref{sl}) and dashed line is for the approximation (\ref{beloborodov}).
Right: the relative error $\delta g_\pm(r,\theta_\mathrm{o})$ of the 
extremal values of $g$-factor. The solid lines are for 
eq.~(\ref{bao_mm}) and the dashed lines are for eq.~(\ref{sl_mm}). 
Upper lines show the relative error of $g_+$, lower lines are for
the relative error of $g_-$. The relative error for the approximation 
(\ref{beloborodov_mm}) is better than $0.1\%$ and thus it is not plotted here.}
\label{fig2}
\end{center}
\end{figure*}

\begin{figure*}
\begin{center}
\includegraphics[width=0.49\textwidth]{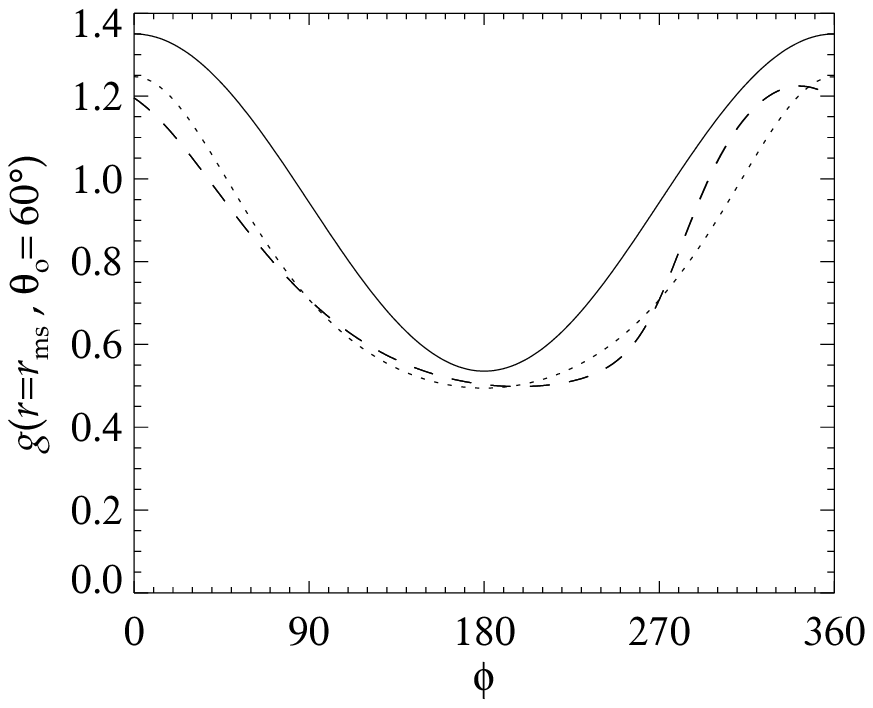}
\includegraphics[width=0.49\textwidth]{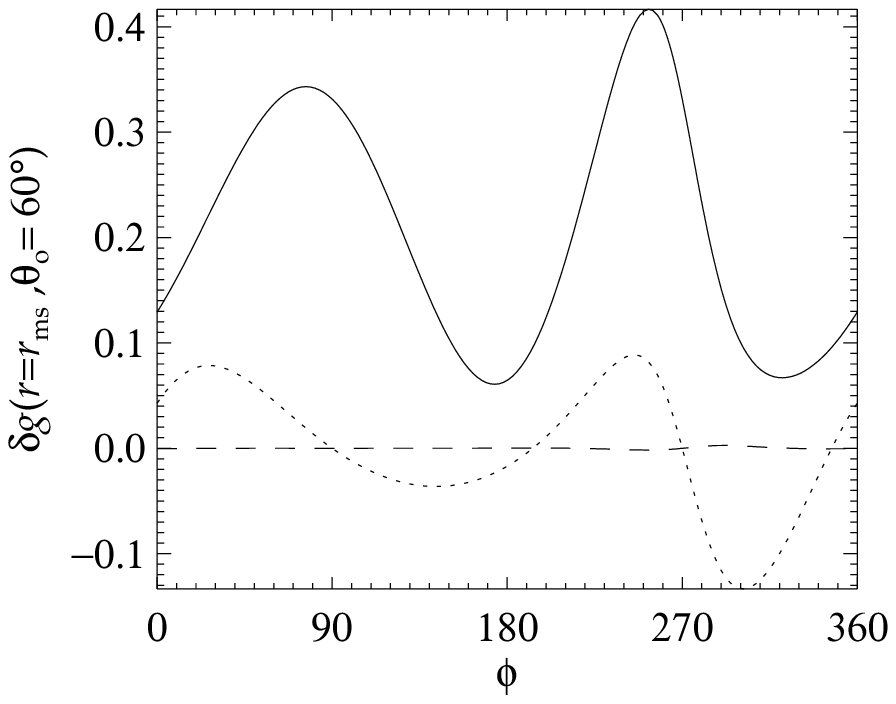}
\caption{Left: $g$-factor as a function of the azimuthal coordinate
$\phi$  for $r=r_{\rm ms}$, $\theta_\mathrm{o}=60^\circ$. The solid line
is for eq.~(\ref{bao}), the dotted line is  for eq.~(\ref{sl}) and the
dashed line is for eq.~(\ref{beloborodov}).  The exact numerical
solution is also plotted by the dashed line and cannot be distinguished
from the approximation (\ref{beloborodov}) in this figure. Right:  as on
the left but for the relative error of $\delta g$ for the three cases of
approximation formulae. Notice that the error of the best one,
eq.~(\ref{beloborodov}), is virtually zero.}
\label{fig3}
\end{center}
\end{figure*}

\begin{figure*}
\begin{center}
\includegraphics[width=0.49\textwidth]{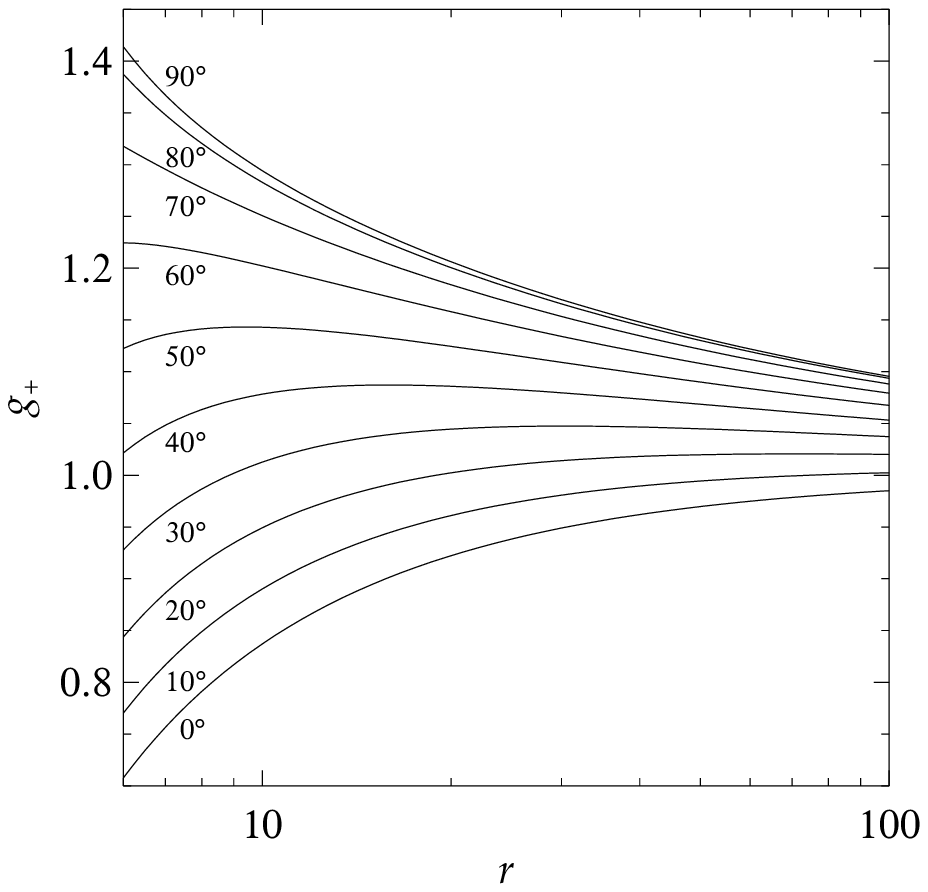}
\includegraphics[width=0.49\textwidth]{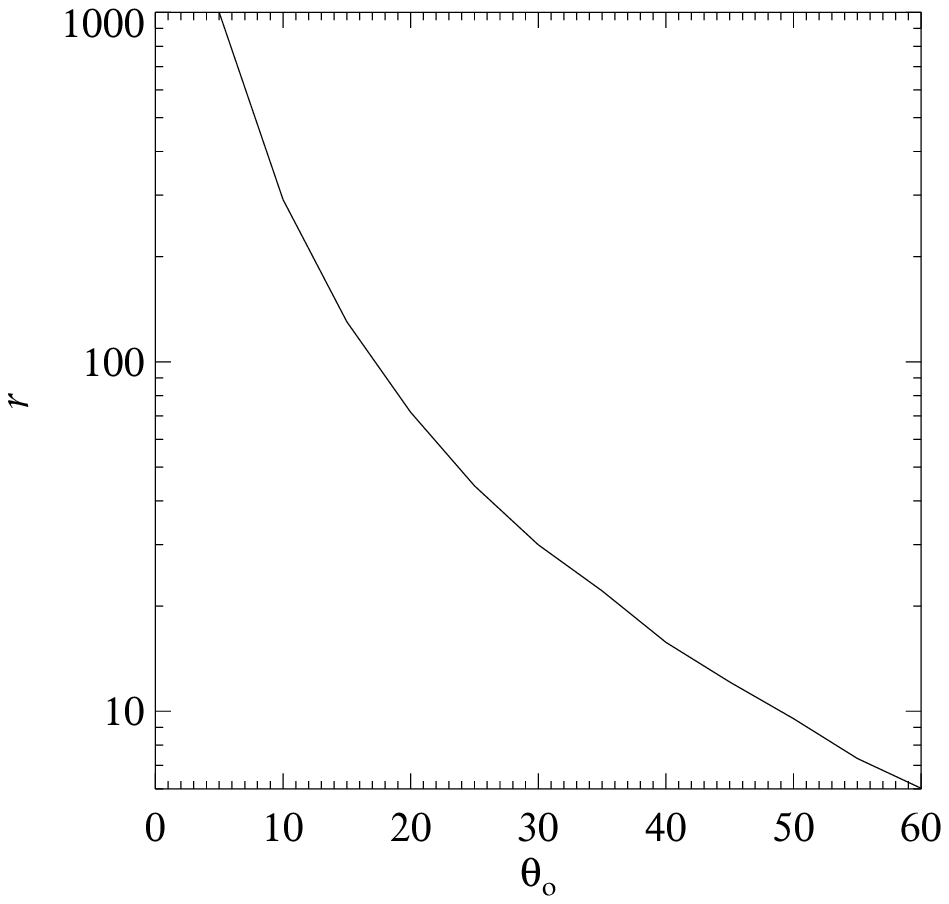}
\caption{Left: the maximum redshift factor $g_+(r)$ as a function of the 
radius. Different curves are parametrized by inclination angle 
$\theta_\mathrm{o}$. Right: the radius at which the function $g_+(r)$
is maximum  as a function of $\theta_\mathrm{o}$.  For
$\theta_\mathrm{o}\gtrsim60^\circ$ the maximum occurs at  marginally
stable orbit where the inner edge of the disc is assumed.}
\label{plotgm}
\end{center}
\end{figure*}

\subsection{Details of the calculation}
\label{sec:details}
In this section we derive the results and provide comparison with
previous  works. The relativistic shift $g$ of the energy of photons, i.e.\
the ratio of the frequency observed by a distant observer to the frequency
emitted in a local frame co-rotating with the disc material, can be
expressed, within the approximation of geometrical optics, in terms of
the four-momentum of photons $p$ and the four-velocity $U$ of the emitting
material and of the distant observer. The shift
$g(r,\phi;\theta_\mathrm{o})$ is a function of polar coordinates in the
disc plane and of observer's inclination,
\begin{equation}
 g(r,\phi;\theta_\mathrm{o})\equiv\frac{E_\mathrm{o}}{E_\mathrm{e}}=
 \frac{p_{\mathrm{o}\,\mu} U^\mu_\mathrm{o}}{p_{\mathrm{e}\,\alpha} U_\mathrm{e}^\alpha}=
 -\frac{p_{\mathrm{o}\,t} }{p_{\mathrm{e}\,\alpha} U_\mathrm{e}^\alpha}\,,
 \label{eq:gfac}
\end{equation}
where $E_\mathrm{o}$ denotes the photon energy measured by a distant
static observer ($r\rightarrow\infty$), $E_\mathrm{e}$ is the locally
emitted energy from a source co-rotating with the disc,
$U_\mathrm{e}^\mu$ is the four-velocity of the disc medium, and
$U_\mathrm{o}^\alpha$ is the observer's four-velocity.

Equation (\ref{eq:gfac}) can be written in the following way:
\begin{equation} 
g(r,\phi;\theta_\mathrm{o}) = \frac{1}{U_{\rm e}^t(r)\,\big[1 - 
            \Omega(r)\,l(r,\phi;\theta_\mathrm{o})\big]},
\end{equation} 
where $U_{\rm e}^t(r)=(1-3/r)^{-1/2}$ is a component of the disc
Keplerian four-velocity, $\Omega(r)=r^{-3/2}$ is the angular velocity and
$l(r,\phi;\theta_\mathrm{o})$ is the azimuthal component of the photon's
angular momentum (constant along null geodesics). The dependence
$l(r,\phi;\theta_\mathrm{o})$ on the polar coordinates in the disc can be calculated
analytically with the help of elliptic functions (Chandrasekhar
\cite{cha92}). 
Let us adopt the approximation in terms of the $\phi$-component of the 
photon momentum in a local frame,
\begin{equation} 
\label{eq:l}
l(r,\phi;\theta_\mathrm{o}) = \frac{n^{(\phi)}(r,\phi;\theta_\mathrm{o})\,e_{(t)}^t(r) +
e_{(\phi)}^t(r)}{n^{(\phi)}(r,\phi;\theta_\mathrm{o})\,e_{(t)}^\phi(r) + 
e_{(\phi)}^\phi(r)},
\end{equation} 
where $n^{(a)}=p^{(a)}/p^{(t)}$ is a unit three-vector in the direction
of photon emission. We choose the local frame attached to an observer
with the tetrad $e_{(a)}^\alpha$ (here, we assumed that $p_t=-1$ and
$p_\phi=l$).

The simplest assumption one can impose is that photons are emitted 
``straight'' toward the observer,
\begin{equation}
n^{(\phi)}(r,\phi;\theta_\mathrm{o}) = -\sin{\phi}\sin{\theta_\mathrm{o}}.
\end{equation}
This was adopted by Zhang \& Bao (\cite{zha91}) and in many subsequent
papers, neglecting the effect of light bending. This turns out to 
provide a sufficiently accurate description in many astrophysically relatistic
situations; however, Zhang \& Bao further assume that the photons move
straight in the direction to an observer in the local frame co-moving
with the disc (i.e. the tetrad $e_{(\rm a)}^\mu$ in eq.~(\ref{eq:l})
corresponds to a Keplerian observer and vector $n^{(a)}$ is expressed
in this tetrad). In this way they also neglect the  aberration due to
rotation of the disc, which enlarges the error especially at small
inclination (pole-on view). The resulting formula for the redshift is
\begin{equation}
\label{bao}
g_{\mbox{\tiny ZB}}(r,\phi;\theta_\mathrm{o}) = \left[\frac{r-2}{r\,(r-3)}\right]^{1/2}
              \left[(r-2)^{1/2}-\sin{\phi}\sin{\theta_\mathrm{o}}\right].
\end{equation}
Extremal values occur on a ring $r=\mbox{const}$ 
$\phi\equiv\phi_{\pm}=\mp90^\circ$,
\begin{equation}
\label{bao_mm}
g_{\mbox{\tiny ZB}}^{\pm}(r,\theta_\mathrm{o}) = \left[\frac{r-2}{r\,(r-3)}\right]^{1/2}
              \left[(r-2)^{1/2}\pm\sin{\theta_\mathrm{o}}\right].
\end{equation}
These approximate formulae suffer from excessive inaccuracy in some
regions of the parameter space. We computed these fractional errors and 
illustrate them in three plots. In Figure~\ref{fig1} we plot the
maximum and minimum relative errors of the redshift factor $\delta_\pm
g(r,\theta_\mathrm{o})$ and the relative error of the  extremal values of
$g$-factor $\delta g_\pm(r,\theta_\mathrm{o})$ on a ring $r=r_{\rm ms}$
(marginally stable orbit). Figure~\ref{fig2} shows the
dependence on radius for $\theta_\mathrm{o}=60^\circ$. Figure~\ref{fig3}
shows the azimuthal dependence of  $g$-factor itself and its relative error.

The main source of error is not the neglect of light bending and, indeed,
considerably better results are achieved if we perform this calculation in the 
local frame of a static observer, still neglecting the light bending. In this 
approximation the tetrad of a static observer is used 
in eq.~(\ref{eq:l}) and the redshift function is 
\begin{equation}
\label{sl}
g_{\rm sl}(r,\phi;\theta_\mathrm{o}) = \left[\frac{(r-2)(r-3)}{r}\right]^{1/2}
                       \frac{1}{(r-2)^{1/2}+\sin{\phi}\sin{\theta_\mathrm{o}}},
\end{equation}
and its extremal values on a ring, which again occur at $\phi_{\pm}=\mp90^\circ$,
are
\begin{equation}
\label{sl_mm}
g_{\rm sl}^{\pm}(r,\theta_\mathrm{o}) = \left[\frac{(r-2)(r-3)}{r}\right]^{1/2}
                                \frac{1}{(r-2)^{1/2}\mp\sin{\theta_\mathrm{o}}}.
\end{equation}
This approach is very successful at estimating the extremal values 
(see the left panel in bottom row of Figure~\ref{fig1}) but the relative 
errors of the redshift factor itself 
remain non-negligible.

Finally, the approach can be refined by assuming that photon trajectories
are curved by gravity and adopting a suitable approximation to their
shapes (Beloborodov's \cite{bel02}). Then the azimuthal component of the
directional three-vector $n^{(a)}$ is (see Appendix~\ref{appendix})
\begin{equation}
n^{(\phi)}(r,\phi;\theta_\mathrm{o}) = -\left[\frac{1-\cos^2{\alpha}}{1-\cos^2{\psi}}\right]^{1/2}
             \sin{\phi}\sin{\theta_\mathrm{o}},
\end{equation}
with $\cos{\alpha} = 1-(1-\cos{\psi})(1-2/r)$ and
$\cos{\psi}=\cos{\phi}\sin{\theta_\mathrm{o}}$. The angle $\alpha$ is
the angle that local static observer measures between the radial
direction and the  direction of emission. The redshift factor is
\begin{eqnarray}
\nonumber
\lefteqn{g_{\mbox{\tiny B}}(r,\phi;\theta_\mathrm{o}) =} \\
\lefteqn{\hspace*{2em} = \frac{\big[r\,(r-3)\big]^{1/2}}{r+\left[r-2+
            4\left(1+\cos{\phi}\sin{\theta_\mathrm{o}}\right)^{-1}\right]^{1/2}
            \sin{\phi}\sin{\theta_\mathrm{o}}}.}
\label{beloborodov}
\end{eqnarray}
Expressions for the extremal values in this case are summarized in the
previous section, see eqs.\ (\ref{beloborodov_mm})--(\ref{f3}).

\begin{table*}[t]
\begin{center}
\begin{tabular}{l|ccccl}
\hline
& ~ & ~ & ~ & ~ & \cr
~~~~~~Source & $E_{\rm{}line}$~[keV] & $r~[GM/c^2]$ & $\theta_\mathrm{o}$~[deg] & $r_{\theta_\mathrm{o}={\rm{}const}}~[GM/c^2]$ & ~~~~~~~Reference \cr
& ~ & ~ & ~ & ~ & \cr
\hline
NGC~3516 & 5.57 & 6--12 & 0--23 & 6.6 $(\theta_\mathrm{o}=20^{\circ})$ & Turner et al.\ (\cite{tur02}) \cr
 & 6.22 & 6--50 & 0--35 & 12.4 $(\theta_\mathrm{o}=20^{\circ})$ & \cr
 & 6.53 & $>$6 & 20--40 & 56/93 $(\theta_\mathrm{o}=20^{\circ})$ & \cr
  \hline
ESO~198-G024 & 5.70 & 6--14 & 0--26 & 7.2 $(\theta_\mathrm{o}=20^{\circ})$ & Guainazzi (\cite{gua03}) \cr
 & 5.96 & 6--21 & 0--30 & 8.9 $(\theta_\mathrm{o}=20^{\circ})$ & Dov\v{c}iak et al.\ (\cite{dov04}) \cr
  \hline
NGC~7314 & 5.84 & 6--17 & 0--28 & 6.3 $(\theta_0=27^{\circ})$  & Yaqoob et al.\ (\cite{yaq03}) \cr
 & 6.61 & $>$6 & 27--41 & 20/93 $(\theta_0=27^{\circ})$  &  \cr
  \hline
Mrk~766 & 5.60 & 6--12.5 & 0--24 & 6.7 $(\theta_\mathrm{o}=20^{\circ})$ & Turner et al.\ (\cite{tur04}) \cr
 & 5.75 & 6--15 & 0--27 & 7.4 $(\theta_\mathrm{o}=20^{\circ})$ &  \cr
   \hline
ESO~113-G010 & 5.40 & 6--10 & 0--20 & 6 $(\theta_\mathrm{o}=20^{\circ})$ & Porquet et al.\ (\cite{por04}) \cr
   \hline 
\end{tabular}
\caption{The narrow features detected so far in AGN which can be interpreted
as the blue horns of a $6.4$~keV iron line arising from an orbiting spot.}
\label{features}
\end{center}
\end{table*}

\section{Comparison with observations}
\label{sec:results}
{\it{}Chandra} and XMM-{\it{}Newton} have
discovered narrow and sometimes transient features in the $5$ to
$7$~keV energy range. Dov\v{c}iak et al.\ (\cite{dov04}) argued that
these lines may be the blue horn of an iron line profile from an
emitting annulus in the accretion disc, a situation which could arise if
an X-ray flare just above the disc survives a non-negligible 
fraction of the orbital period or more.

These features offer the possibility to measure the black hole
mass (see discussion in Dov\v{c}iak et al.\ \cite{dov04}), and indeed
Iwasawa et al.\ (\cite{iwa04}) applied this method to the Seyfert~1
galaxy, NGC~3516. The quality of data, however, is in most cases
insufficient for this purpose. For the time being, and waiting for the
next generation of X-ray satellites like Constellation-X and XEUS, we
estimate the orbiting spot parameters.

In Figure~\ref{plotgm} (left panel) we show the maximum energy shift
$g_+(r)$, which corresponds to the blue edge of the line and the peak
flux when lensing is neglected or unimportant. Only direct image photons
were considered (higher-order images can be important only for
equatorial inclinations of $\theta_\mathrm{o}\sim90^{\circ}$; see below).
In this case the energy shift  is given by eq.~(\ref{beloborodov_mm}). 
We checked that the exact value (based on the full geodesic equation and
the ray tracing method) for the energy shift is indistinguishable from
the approximation adopted in this figure. This plot permits
determination of ranges of allowed ($r,\theta_{\rm{}o}$) pairs from the
measured energy of the narrow feature. A complementary plot is given in
the right panel where we show radius at which the maximum energy shift
occurs for a given value of inclination angle
$\theta=\theta_\mathrm{o}$. 

One can ask whether the lensing effect, which we have neglected in our 
present discussion, could change the conclusion about the observed
energy  of the Doppler boosted horn of a relativistic line.  The reason
is that for higher inclination angles  ($\theta_{\rm o}\gtrsim60^\circ$)
the lensing effect and bending of  light rays (that changes the
projected area of the spot) can play an important role: thus the
dominating feature need not be a blue horn associated with the
maximum shift of energy but, instead, a new peak of the maximum
observed flux that should arise redwards of $g_+(r)$ (see e.g.\ Matt et
al.\ \cite{mat93}). We have therefore employed a ray-tracing code
(Dov\v{c}iak et al.\ \cite{dov04b})  taking all these general-relativity
effects into  account, and examined time-dependent model spectra. Indeed
a new peak is  formed by photons arriving from another azimuth
(different from $-90^{\circ}$) of the spot orbit. The secondary maximum
is in the point where the effect of energy shift, lensing, projection
area and time delay combine to produce the maximum
amplification. However, the time delay, and especially the function 
$g(r,\phi)$, vary rapidly across the region of the disc most affected
by lensing.  As a result, the new peak  becomes spread over a broad range
of energy and does not form a narrow line. In other words, even if the
lensed feature may have a large equivalent width and can contribute
significantly to the total photometric flux, this effect is not relevant
for our present discussion of the narrow line origin,  as the resulting
feature is not narrow at all.

In Table~\ref{features} we summarize the values of $r$ and $\theta_{\rm
o}$ for the most convincing observational cases known.
For all  features a solution in which the orbiting
spot is in the innermost stable orbit for a Schwarzschild black hole,
i.e $r=6$, is acceptable. The inclination angle is generally expected to
be low or moderate (the inner accretion disc is presumably aligned with
the outer torus).  If the features we discuss originate at or near the
innermost stable orbit, the corresponding inclination angle is the upper
value in the table. Naturally, as this angle must be the same for
different features of the same source, the actual allowed range is the
intersection of the various ranges: $20$--$23^{\circ}$ (NGC~3516),
$0$--$26^{\circ}$ (ESO~198-G024), $27$--$28^{\circ}$ (NGC~7314),
$0$--$24^{\circ}$ (Mrk~766). Assuming, therefore, the same inclination
angle for a given source, and assuming this angle to be $20^{\circ}$ for
all sources but one, the corresponding radii are listed in the fourth
column of the table. For NGC~7314, if the two features are both
the blue horn of an annular emission, then the inclination angle is
constrained to be between $27^{\circ}$ and $28^{\circ}$;  for this
source an angle of $27^{\circ}$ has been assumed. 
For the only two features that are blueward of the line rest
energy, the resulting radii could not be determined uniquely. This is
explained by the fact that the line of $g={\rm{}const}$ can intersect
the curve $g_+(r)$ at two different radii (see Fig.~\ref{plotgm}).

\begin{figure*}
\begin{center}
\hfill~
\includegraphics[width=0.3\textwidth]{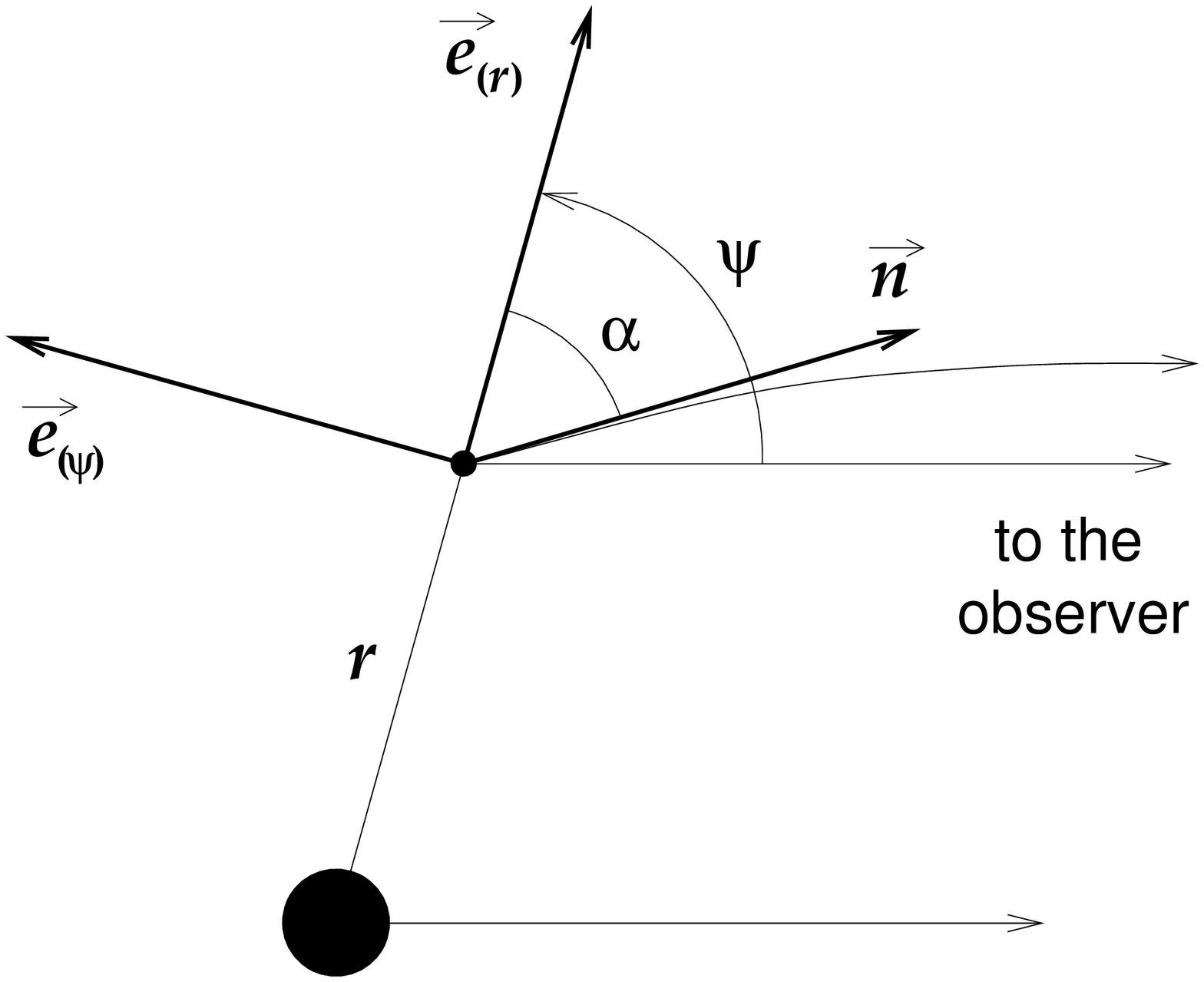}
\hfill~
\includegraphics[width=0.3\textwidth]{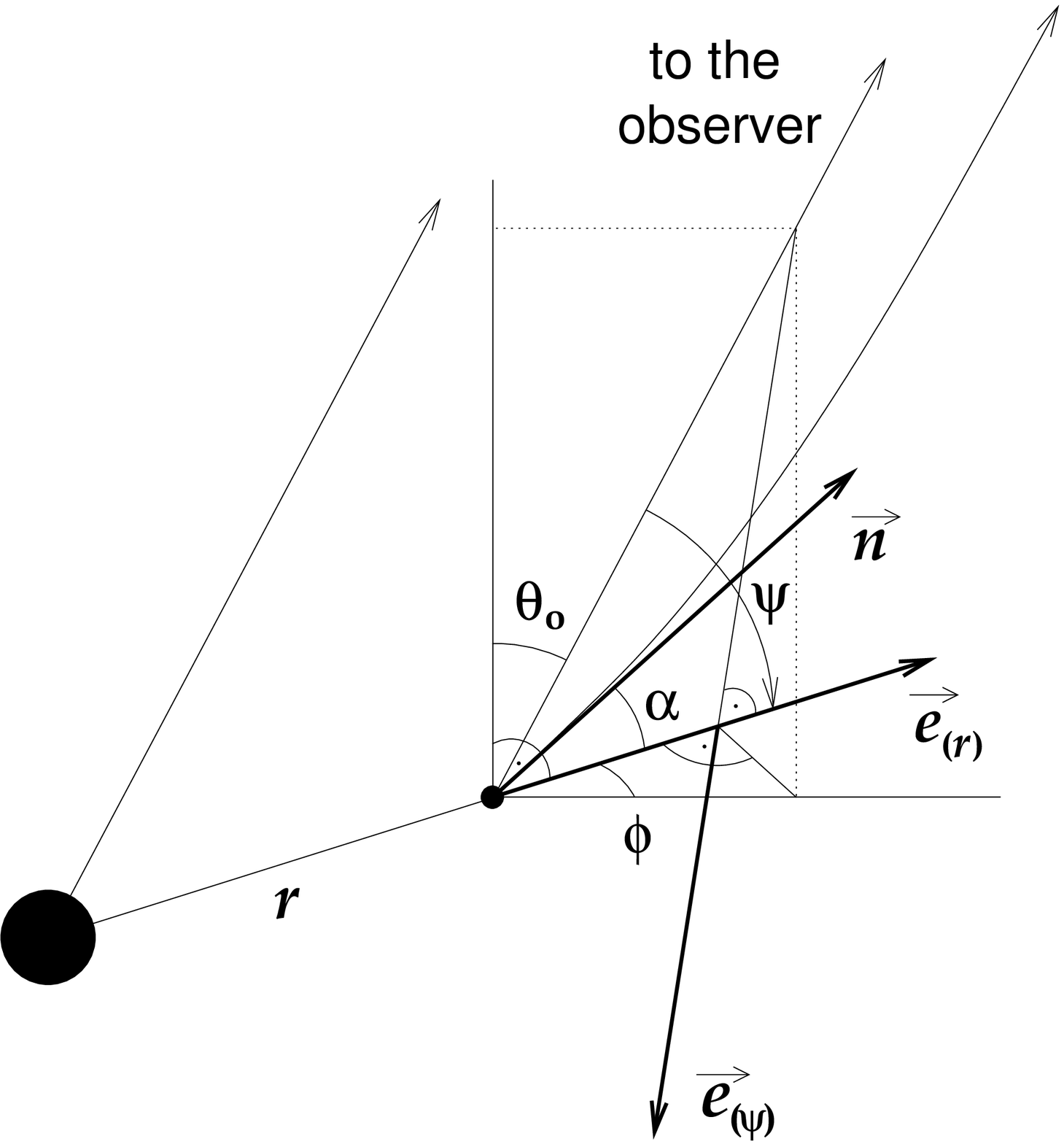}
\hfill~
\caption{A schematic representation of the plane of photon motion
(left), and the definition of angles and directions in Schwarzschild 
coordinates $(r,\theta,\phi)$ (right). The direction vector $\vec{n}$ is
tangent to the photon ray projected into three-space.}
\label{fig5}
\end{center}
\end{figure*}

The values of the parameters in the table have been derived assuming
a rest-frame energy of the line $E_0\sim6.4$~keV, as appropriate for
neutral matter. However, the iron may be ionized and the corresponding
$E_0$ higher. The maximum possible energy is for H--like iron, and is
$E_0\sim6.96$~keV (the line is a doublet again). Because in this case the
redshift has to be larger, the upper limits to the radius and the
inclination angle are reduced. For instance, for the reddest feature
(that of ESO~113-G010) the allowed ranges are now $6\lesssim{r}\lesssim7.6$ for the
dimensionless radius and $\theta_{\rm{}o}\lesssim11^{\circ}$ for the inclination.

As discussed above, for high inclination angles the maximum flux could
occur not in correspondence with the blue peak, as we assumed in this
paper, but instead with the moment of the spot crossing a caustic at an
intermediate phase. In this case the resulting inclination comes out
larger. We have explored and rejected this lensing effect as the origin
of narrow lines not only due to the large inclination, but also because
the width of the resulting spectral feature exceeds substantially the
width of the blue peak.  In no case do the observed features require a
spinning black hole, as a solution with $r>6$ is always found.

\section{Discussion}
Variability properties contain a wealth of information on accreting black
holes, but much of it is lost in the process of time integration in
the detector. Several ways have been proposed to measure the
parameters of accreting black holes, which rely on the source
variability: for example, limits on the black-hole mass and its rate of
rotation can be derived from time delays between variations of the
emission-line strength and of the continuum, and from temporal changes
of the observed emission lines. Time-dependent spectra have not yet
provided the anticipated breakthrough in our understanding; they
still remain a challenge for theorists and observers alike. This is
caused by  difficulties and non-uniquenesses in the interpretation of
the available dynamical spectra. On the side of observation a trade-off
must be accepted between energy and temporal resolution which also
compromises the usefulness of timing spectroscopy. That is why we search
for sharp features, and the narrow redshifted lines are promising.

Here we showed that a simple and accurate formula can be given
for the estimate of the energy shift. The analytical formula circumvents the
necessity of numerical integration of photon paths and can be
straightforwardly used to determine the orbiting spot parameters from the
measured energy of the narrow feature, with an accuracy that is more
than sufficient. 
The use of this formula implies the assumptions that the 
feature corresponds to the blue peak in profile, that the effect 
of black-hole rotation is negligible and that the disc is 
essentially planar.

\begin{acknowledgements} 
The authors of this paper have been supported by the Czech Science 
Foundation under grants 205/03/H144 (TP), 205/05/P525 (MD) and
205/03/0902 (VK), and by Italian Ministry of Research under grant {\sc
cofin-03-02-23} (GM). Financial support from the Academy of Sciences of
the Czech Republic is also gratefully acknowledged (grant
IAA\,300030510). The Astronomical Institute has been operated under the
project AV0Z10030501.
\end{acknowledgements}

\appendix
\section{Direction of photon emission}
\label{appendix}
In the Schwarzschild metric a ray can be defined by three points: the
origin of the coordinate system (which coincides with the center of the
black hole), location of the spot (in the disc plane), and the observer 
(at radial infinity). See Figure~\ref{fig5} for the geometrical setup. 
Then the directional vector $n^{(a)}$ of a
photon can be expressed with respect to the local static frame. We have
employed Beloborodov's (\cite{bel02})  approximation of light rays. One
can write
\begin{equation}
n^\mu = \cos{\alpha}\,e_{(r)}^\mu - (1-\cos^2{\!\alpha})^{1/2}e_{(\psi)}^\mu,
\end{equation}
where the unit vector $e_{(\psi)}^\mu$ is perpendicular to the tetrad vector
$e_{(r)}^\mu$ and lies in the plane of the photon's motion. Thus 
(see the right panel in Figure~\ref{fig5})
\begin{equation}
e_{(\psi)}^\mu = r^{-1}(1-\cos^2{\!\psi})^{-1/2}(0, 0, \cos{\theta_\mathrm{o}}, 
 \sin{\phi}\sin{\theta_\mathrm{o}})
\end{equation}
with $\cos{\psi}=\cos{\phi}\sin{\theta_\mathrm{o}}$.
The components of the directional vector $n^{(a)}=n^\mu e_{(a)\mu}$ in a local
static frame are
\begin{eqnarray}
n^{(r)} & = & \cos{\alpha}, \\
n^{(\theta)} & = & -\left[\frac{1-\cos^2{\!\alpha}}{1-\cos^2{\!\psi}}\right]^{1/2}
                \cos{\theta_\mathrm{o}}, \\
n^{(\phi)} & = & -\left[\frac{1-\cos^2{\!\alpha}}{1-\cos^2{\!\psi}}\right]^{1/2}
               \sin{\phi}\sin{\theta_\mathrm{o},}.
\end{eqnarray}
Here, we have used the tetrad of a static observer located in the equatorial
plane:
\begin{eqnarray}
e_{(t)}^\mu & = & (1-2/r)^{-1/2}(1, 0, 0, 0), \\ 
e_{(r)}^\mu & = & (1-2/r)^{1/2}(0, 1, 0, 0), \\
e_{(\theta)}^\mu & = & r^{-1}(0, 0, 1, 0), \\
e_{(\phi)}^\mu & = & r^{-1}(0, 0, 0, 1).
\end{eqnarray}

\end{document}